\documentclass[useAMS,usenatbib]{mn2e}
\usepackage{graphicx}
\usepackage{amssymb}
\usepackage{lscape}
\usepackage{ulem}
\usepackage{txfonts}

\def\kms{km~s$^{-1}$}
\def\cm{cm$^{-2}$}
\def\lya{Ly$\alpha$}
\def\nhi{$N$(H\,{\sc i})}
\def\hi{H\,{\sc i}}
\def\si2{Si\,{\sc ii}}
\def\s2{S\,{\sc ii}}
\def\mg2{Mg\,{\sc ii}}
\def\fe2{Fe\,{\sc ii}}
\def\chr2{Cr\,{\sc ii}}
\def\zn2{Zn\,{\sc ii}}
\def\mn2{Mn\,{\sc ii}}
\def\c2s{C\,{\sc ii}$^{\star}$}

\def\ion#1#2{{\rm #1}\,{\sc #2}}

\title[Ionization corrections in a multi-phase ISM.]{Ionization 
corrections in a multi-phase interstellar
medium: Lessons from a $z_{\rm abs} \sim 2$ sub-DLA.}

\author[Milutinovic et al.] {Nikola Milutinovic$^1$, Sara
  L. Ellison$^1$,
  J. Xavier Prochaska $^2$, Jason Tumlinson $^3$\\
  $^1$Department of Physics and Astronomy, University of Victoria,
  Victoria, B.C., V8P 1A1, Canada\\
  $^2$ Department of Astronomy and Astrophysics, UCO/Lick Observatory, University of California, 1156 High Street, Santa Cruz, CA 95064, USA \\
  $^3$ Space Telescope Science Institute, 3700 San Martin Drive,
  Baltimore, MD 21218, USA }

\begin{document}

\maketitle

\begin{abstract}

  We present a high resolution (FWHM$=2.7$~km~s$^{-1}$), high S/N
  echelle spectrum for the $z_{\rm em} = 2.26 $ QSO J2123$-$0050 and
  determine elemental abundances for the $z_{\rm abs} = 2.06$ sub-DLA
  in its line of sight.  This high redshift sub-DLA has a complex
  kinematic structure and harbours detections of neutral (\ion{S}{i},
  \ion{C}{i}), singly (e.g. \ion{C}{ii}, \ion{S}{ii}) and multiply
  ionized (e.g. \ion{C}{iv}, \ion{Si}{iv}) species as well as
  molecular H$_2$ and HD. The plethora of detected transitions in
  various ionization stages is indicative of a complex multi-phase
  structure present in this high redshift galaxy.  We demonstrate that
  the ionization corrections in this sub-DLA are significant (up to $\sim$
  0.7 dex). For example, if no ionization correction is applied, a super-solar
  metallicity is derived ([S/H] = $+0.36$), whereas a single phase
  ionization correction reduces this to [S/H] = $-0.19$.  The
  theoretical impact of a multi-phase medium is investigated through
  Cloudy modelling and it is found that the abundances of Si, S and Fe
  are always over-estimated (by up to 0.15 dex in our experiments) if
  a single-phase is assumed.  Therefore, although Cloudy models
  improve estimates of metal column densities, the simplification of a
  single phase medium leaves a systematic error in the result, so that
  even ionization-corrected abundances may still be too high.
  Without ionization corrections the properties of this
  sub-DLA appear to require extreme scenarios of nucleosynthetic
  origins.  After ionization corrections are applied the ISM of this
  galaxy appears to be similar to some of the sightlines through the
  Milky Way.

\end{abstract}


\section{Introduction}

The measurement of rest-frame ultra-violet resonance lines in damped
Lyman alpha (DLA) systems is currently the most successful technique
for determining chemical abundances at high redshift (e.g. Prochaska
et al. 2003; Wolfe, Gawiser \& Prochaska 2005).  The largest
compilations of DLA abundances include column densities for over a
dozen different elements in some 200 absorbing galaxies
(e.g. Prochaska et al. 2007b; Dessauges-Zavadsky et al. 2009).  
It is usually assumed that,
due to the high \nhi\ column density of the absorber, ionization
corrections are negligible and that total elemental column densities
can be derived from the dominant ionization state, i.e. the lowest
energy species where the ionization potential is above 13.6 eV.  For
the majority of elements observed in absorption, such as silicon,
iron, zinc, sulphur, nickel, titanium and chromium, this is the singly
ionized species, leading to the approximation N({\rm X}) $\approx$
N(\ion{X}{ii}).  Exceptions include oxygen and nitrogen whose first
ionization potential is close to that of hydrogen and whose ionization
balance is governed by charge exchange so that the neutral atom is the
dominant species.  Therefore, although absorption from non-dominant
species is observed in DLAs (such as \ion{C}{iv}, \ion{Mg}{i} and
\ion{Na}{i}) their contributions are minor and not usually considered
in abundance determinations.  The robustness of this approach to
abundance calculations has been tested numerous times in the
literature (Viegas 1995; Howk \& Sembach 1999; Vladilo et al 2001;
Prochaska et al. 2002a), usually through models that employ the
ionization code Cloudy (Ferland et al. 1998). Although the exact
magnitude of the corrections depends on a variety of input parameters,
most notably the shape and normalization of the ionizing background
(e.g. Viegas 1995; Howk \& Semback 1999), in general the literature
agrees that the majority of DLA column densities do not require
significant ionization corrections (e.g.  Vladilo et al. 2001).
However, there are a few noteworthy cases where ionization corrections
may be important (Prochaska et al. 2002a; Prochaska et al. 2002b;
Dessauges-Zavadsky et al. 2004; Dessauges-Zavadsky et al. 2006;
Ellison et al. 2010).

As the column density of neutral hydrogen decreases, the
self-shielding approximation becomes less robust as more ionizing
photons penetrate the cloud.  For example, it has been shown that the
ratio of \ion{Al}{iii}/\ion{Al}{ii} (or some proxy for \ion{Al}{ii} if
it is saturated) tends to increase as the \nhi\ decreases (Vladilo et
al. 2001).  As studies of quasar absorption line systems began to push
down the \nhi\ scale to investigate the nature of sub-DLAs\footnote{Different
naming conventions have emerged for absorbers that exhibit damping wings,
yet do not qualify as DLAs. The most common alternative name
is super-Lyman limit system, e.g. O'Meara et al. (2007).} with 19.0
$<$ log \nhi\ $<$ 20.3 \cm\ (e.g. Peroux et al.  2003) it was natural
to re-assess the need for ionization corrections (Dessauges-Zavadsky
et al. 2003; Meiring et al. 2007).  Once again, it was found that, in
general, ionization corrections were small (below 0.2 dex) and
observed column densities are therefore usually converted directly
into metallicities (e.g. Peroux et al.  2007; Dessauges-Zavadsky et
al. 2009). Nonetheless, for some sub-DLAs, ionization corrections
appear to be non-negligible (e.g. Richter et al. 2005; Quast et
al. 2008).

\subsection{This work: methodology and context}

In this paper we present the case of a low column density sub-DLA
towards the QSO J2123$-$0050 at $z_{\rm abs} = 2.06$ whose unusual
properties led us to re-assess the issue of ionization corrections for
sub-DLAs.  Since the methodology of this work differs from previous
studies of ionization corrections in sub-DLAs, it is useful to
summarise our approach at the outset.  The superb data quality, in
terms of both resolution ($R \sim$ 100,000) and S/N (up to 50 per
pixel), has permitted a rare chance to demonstrate the presence of a
multi-phase medium, with contributions from both a mostly neutral
component, and a more highly ionized component.  Although many works
on sub-DLAs have previously investigated ionization corrections in
sub-DLAs (see above references), these investigations have uniformly
assumed that the gas measured in absorption originates from a single
phase.  Since the sightline towards J2123$-$0050 firmly establishes
that this is not always the case, our main objective is to answer the
following question: `If multi-phase media are common in (sub)-DLAs,
what will be the effect of calculating Cloudy ionization corrections
with only a single phase?'. 

 In order to address this question we create a range of models with
two-phase media, in which we vary the fractions of the `cold' and
`warm' gas (as characterised by different ionization parameters).
These models yield column densities of species such as SiII, FeII,
AlII and AlIII.  The key to our model philosophy is that we take these
column densities as `observed' values and adopt the usual empirical
strategies of observers, and the assumption that the medium is a
single phase, to attempt to derive the input parameters.  In essence,
we are attempting to recover the input model, but using an incorrect
assumption about the ionization structure.  Ultimately, we will
quantify how wrong the derived elemental abundances will be under the
assumption of a single phase ISM.  The utility of this approach is
that the relative contributions of components of real multi-phase
absorbers can rarely be constrained.  Multi-phase models are therefore
not usually possible in practice.  Our models therefore provide an
indication of the likely error associated with the (necessary)
single-phase approach.  Additional uncertainties, such as those
associated with geometry or in the atomic data, are not considered in
this investigation.

The paper is laid out as follows.  In Section \ref{obs_sec} we
describe the observations and data reduction of the QSO J2123$-$0050
and describe the measurement of the \nhi\ and metal species column
densities.  In Section \ref{abund_sec} we discuss the puzzling nature
of this sub-DLA in the context of its ISM properties, if no ionization
corrections are applied.  The general fidelity of single phase models
in the multi-phase case is quantified in Section \ref{cloudy_sec}
and evidence for a multi-phase medium in J2123$-$0050 discussed
in Section \ref{mp_sec}.
Returning to the specific case of J2123$-$0050, photoionization
corrections are calculated and applied in Section
\ref{sec_corrections}. Results are summarised in Section
\ref{summary_sec}.  The molecular content (H$_2$ and HD) of this
absorber are investigated in 2 companion papers (Malec et al. 2010;
Tumlinson et al. in preparation).

\section{Observations and Data Reduction}\label{obs_sec}

The sub-DLA towards J2123$-$0050 was observed with the High Resolution
Echelle Spectrograph (HIRES) as part of a program to follow-up
metal-strong absorbers identified in the Sloan Digital Sky Survey
(SDSS) by Herbert-Fort et al. (2006).  These absorbers are selected in
an automated fashion which searches for absorption in $14$ prominent
resonance lines. From the $\sim20000$ quasars searched, more than
$2000$ metal absorption systems have been identified.  In $>95\%$ of
cases where \lya\ is covered in the SDSS spectrum, the \hi\ column
density is found to be large and the absorber would be classed as a
DLA (Kaplan et al. 2010).  Follow-up of a sub-sample of these metal-line 
selected DLAs with
HIRES has shown that the metallicity of these DLAs is approaching the
solar value, even at $z \sim 2$ (Kaplan et al. 2010). The main results
of our high resolution survey will be presented elsewhere.  In this
paper, we focus on the case of the $z_{\rm abs} \sim 2.06$ absorber
towards J2123$-$0050 ($z_{\rm em} = 2.26$, $r$ = 16.44).

A preliminary HIRES exposure of 2 $\times$ 2700 seconds of
J2123$-$0050 with the C1 decker (equivalent to a 0.86 arcsec wide
slit) obtained on August 18th 2006, revealed the detection of H$_2$
and fine structure lines of carbon at $z = 2.05930$. Given the good
observing conditions and the brightness of the QSO, further HIRES
observations were obtained on April 19 2006 with a total exposure time
of $10,800$~s but with the E1 decker ($0.4$ arcsec wide), which yields
a resolution of $R \sim 100,000$ (FWHM$\sim3$~{\kms}).  Such high
resolution is beneficial when studying the coldest phases of the ISM,
allowing us to potentially resolve even very narrow components of the
diffuse gas (e.g.  Narayanan et al. 2006).  All observations were
conducted with the blue cross disperser, an echelle angle of
0$^\circ$, and cross-disperser angle of 1.0275$^\circ$, yielding a
total wavelength coverage of approximately 3000 -- 6000 \AA.  For
calibration purposes, a set of standard trace flats were obtained, as
well as the spectra of ThAr lamps (arcs) using the same instrument
settings. We also obtained a set of pixel flats (lamp flats) at the
beginning of the observing run to determine the pixel-to-pixel
variation across the detectors.

The data were reduced using the HIRedux routine which is a part of the
XIDL package\footnote{XIDL is publicly available at
http://www.ucolick.org/$\sim$xavier/IDL/index.html}.  The reduction
involved the following procedures:

\begin{enumerate}

\item A flat field frame was produced from a stack of approximately
30 flat field exposures for each of the detector's chips. 

\item In a similar manner, a combined trace flat frame is produced as
a median over the series of standard flat images taken during the
observing night.  This frame is then used to define (trace) the
echelle order boundaries (and find the order of curvature), and to
determine the slit profile, which is used to correct the illumination
pattern of the science frames. 

\item A wavelength solution is derived from the spectra of a ThAr lamp
taken with the same setup as in the science frames. HIRedux performs a
1D wavelength solution by fitting a low-order Legendre polynomial to
the pixel values in the ThAr exposures versus the laboratory
wavelengths along the spatial centres of each order.  The code then
performs a 2D fit to all the lines from the 1D solution. Finally,
the pipeline derives the 2D wavelength map giving both the wavelength
solution, and the line tilts for all orders over the full echelle
footprint.  

\item After the raw science frame images are flattened and cosmic rays
are flagged, the final step in the reduction process is the extraction
of the object and the sky spectra.   The sky background is estimated
from the pixels that fall well beyond the object aperture, taking into
account diffuse scattered light, which is estimated by interpolating
the pixel counts in the gaps between the pixel orders. After this, the
procedure derives the spatial profile of the object point spread
function, and performs an optimal extraction based on the order trace. 

\item The individual exposures were coadded by combining each order
separately, yielding a final, unnormalized 2D spectrum.

\item The continuum is fitted manually using the XIDL routine {\it
x\_continuum}.  The routine allows the user to select the parts of the
continuum unaffected by absorption and then performs a minimum
$\chi^2$ fit on the selected data points using a spline function of a
given order (for the HIRES data presented here the usual value of the
spline order is around 8).

\item The echelle orders are combined after normalization into a 1D
spectrum.  In regions of order overlap, the orders are averaged,
weighting by the square of median signal-to-noise ratio. The final
spectrum has a S/N of $\sim15$ per pixel at $3100$~{\AA}, $\sim30$ at
$3500$~{\AA}, and $\sim40$ at $5100$~{\AA}.

\end{enumerate}

\subsection{\nhi\ determination}

\begin{figure}
\centering
\includegraphics[width=9cm]{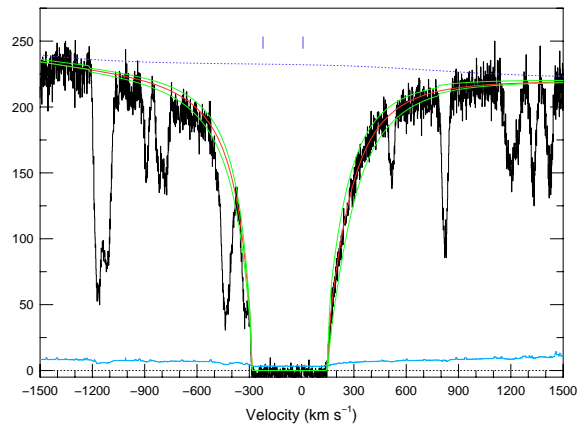}
\caption{Fit to the Ly$\alpha$ line with total
log$N$(\hi)=$19.25{\pm}0.2~$ \cm\ (red line).  The blue dotted line
represents the continuum fit and green lines are the $3\sigma$ bounds
respectively. A two component fit (shown by tick marks) is required to
adequately fit the asymmetric profile. The column densities of
separate components are log$N$(\hi~$z=2.05930$)=$19.18{\pm}0.15~$\cm,
and log$N$(\hi~$z=2.05684$)=$18.40{\pm}0.30~$\cm.  The lower solid
(cyan) line shows the 1 $\sigma$ error array.}\label{fig:Lya}
\end{figure}

To perform the column density measurement of the \lya\ absorption, the
continuum and the line profile were simultaneously fit in a
blaze-corrected section of the spectrum. The fit was performed using
the \textit{x$\_$fitdla} routine of XIDL.  The \lya\ profile is
clearly asymmetric, with the red wing showing stronger damping. The
assymetry indicated that a two-component fit was necessary. The
redshifts of the components were fixed at the redshifts of the
strongest metal line absorption in the two kinematically distinct
metal line complexes detected in the system (see the next
Section). The fit is presented in Figure~\ref{fig:Lya}. The column
densities of the separate components are
log$N$(\ion{H}{i}~$z=2.05930$)=$19.18{\pm}0.15$ \cm, and
log$N$(\ion{H}{i}~$z=2.05684$)=$18.40{\pm}0.30$ \cm, yielding a total
column density of neutral hydrogen of log$N$(\hi)$=19.25\pm0.2$
\cm. Note that these uncertainties are dominated by systematic
(e.g. continuum fitting), not statistical error. 

\subsection{Metal column density measurements}

\begin{figure}
\centerline{\rotatebox{0}{\resizebox{12cm}{!}
{\includegraphics{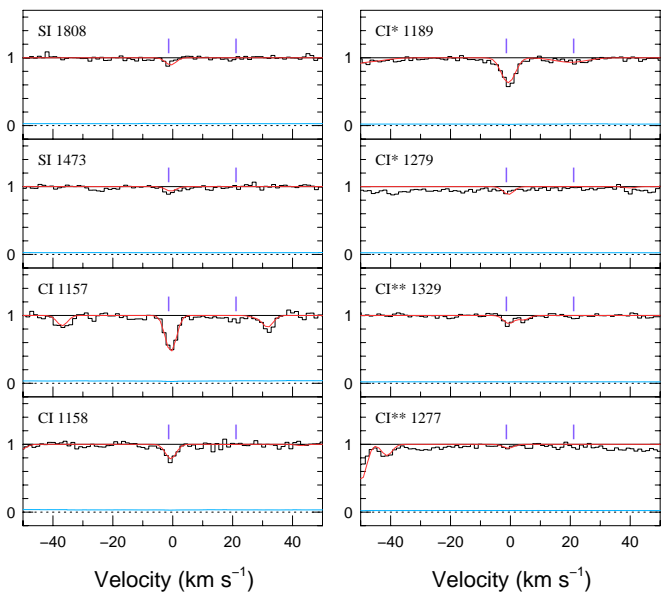}}}}
\caption{\label{fig:CIfig} Fit to neutral carbon and sulphur lines
towards J2123$-$0050 on a velocity scale relative to $z=2.05930$. The
ticks mark the position of the velocity components which give rise to
molecular absorption. The data is presented in black, the error array
is given in blue, and the fit is the red line.}
\end{figure}

\begin{figure}
\centerline{\rotatebox{0}{\resizebox{10cm}{!}
{\includegraphics{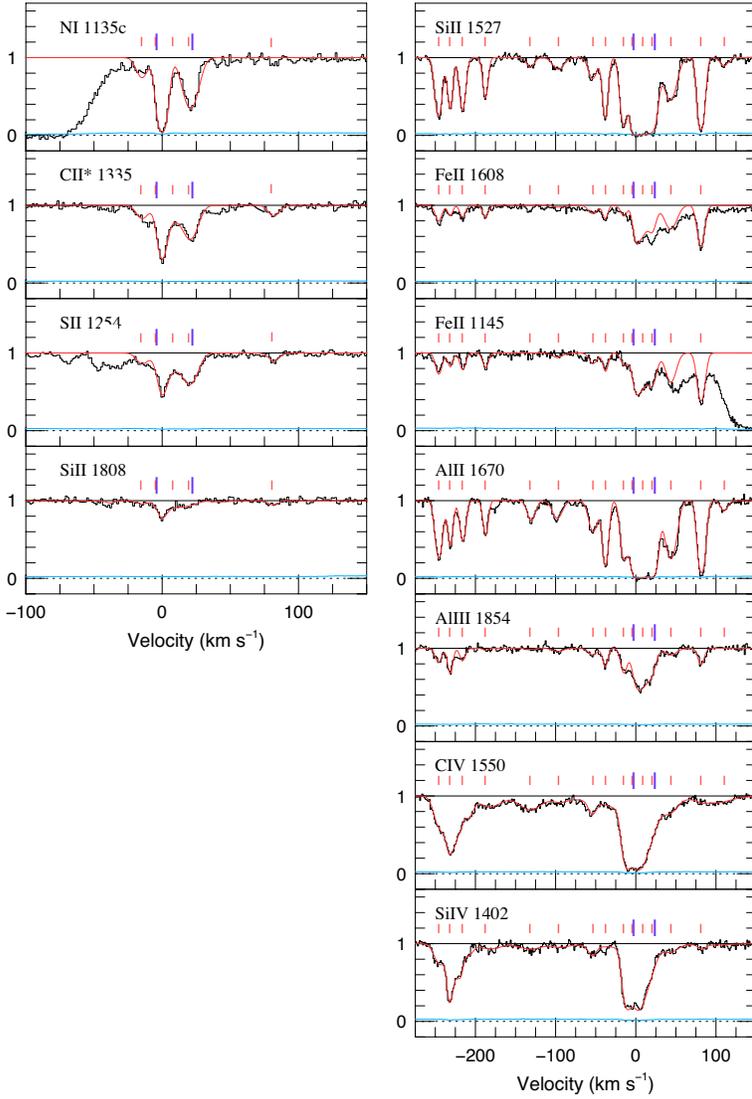}}}}
\caption{\label{fig:metfig} Metal lines towards J2123$-$0050 on a
velocity scale relative to $z=2.05930$.  The long blue ticks mark the
position of the velocity components which give rise to molecular
absorption. The short red tick marks indicate other fitted compoents.
The data is presented in black, the error array is given in blue, and
the fit is the red line.}
\end{figure}

In order to derive the total column density of multi-component metal
line complexes VPFIT
9.3\footnote{www.ast.cam.ac.uk/$\sim$rfc/vpfit.html} was used on the
normalized data. VPFIT is a multiple Voigt profile fitting code that
calculates a maximum likelihood fitting function to the data.  The
code is adapted to fit multiple lines simultaneously, which allows for
efficient identification of blends. The goodness of fit is assessed in
VPFIT by $\chi^2$ statistics. The error estimates of the fitting
parameters also include the uncertainties induced by self-blends, as
well as blends due to unidentified lines. The errors on individual
components are fairly poorly constrained, but the total column density
error can be more accurately quantified and may be quite small,
especially if several species are fit simultaneously with the same
structural model.  Lower limits are reported for species with only
saturated transitions and 3 $\sigma$ upper limits are quoted for
non-detections using the following equations:

\begin{equation}
	W_{obs(3~\sigma)}=3 \times FWHM/(S/N),
\end{equation}
\begin{equation}
	W_r=W_{obs}/(1+z),
\end{equation} 
\begin{equation}\label{cog_eqn}
	N=1.13e20 \times W_r/ (\lambda_r^2 \times f),
\end{equation}

\noindent where $W_r$, and $\lambda_r$ are rest-frame equivalent width and
wavelength, $f$ is oscillator strength, and $(S/N)$ is signal to noise
ratio at the observed line wavelength.  Eqn \ref{cog_eqn} comes from
the linear part of the curve of growth (e.g. Pagel 1997).

\begin{table}
\centering
\caption{Neutral carbon and silicon column densities from Voigt
profile fits.  }
\begin{tabular}{cccccc}
\hline \hline
z & b      & log N(CI) & log N(CI*) & log N(CI**) & log N(SI)  \\
  & (km~s$^{-1}$) & ({cm$^{-2}$})    & ({cm$^{-2}$})     & ({cm$^{-2}$})      & ({cm$^{-2}$})     \\
 
\hline \hline
 2.05930 & 1.8$\pm$0.3  &  13.73$\pm$0.02  &  13.42$\pm$0.03 & 12.43$\pm$0.02  &   12.08$\pm$0.05   \\
 2.05955 & 7.3$\pm$4.0  &  12.91$\pm$0.03  &  12.74$\pm$0.10 & 12.16$\pm$0.11  &         ... \\
\hline    
\multicolumn{2}{l}{Total} 		&  13.79$\pm$0.02  &  13.50$\pm$0.03  & 12.62$\pm$0.03  &   12.08$\pm$0.05  \\ 
\hline \hline
\label{tab:ions_neutral}
\end{tabular} 
\end{table}

\begin{landscape}
\begin{table}
\begin{center}
\begin{tabular}{lcccccccccccccc}
\hline \hline
z         &  b                & log N(SiII)       & log N(FeII)     & log N(NiII)     & log N(AlII)     &  log N(AlIII)        & log N(SII)       & log N(NI)        & log N(CII)     & log N(CII*)        \\
          &  (km~s$^{-1}$)         &  (cm$^{-2}$)        &  (cm$^{-2}$)    &  (cm$^{-2}$)        &  (cm$^{-2}$)        &  (cm$^{-2}$)         &  (cm$^{-2}$)       &  (cm$^{-2}$)    &  (cm$^{-2}$)       & (cm$^{-2}$)          \\
\hline \hline
2.05684  &  5.2 $\pm$ 0.2  &  13.43 $\pm$ 0.02 & 12.93 $\pm$ 0.06   &   ...    &   ...       &  11.86 $\pm$ 0.04   &      ...         &       ...        &      ...        &         ...        \\
2.05698  &  3.9 $\pm$ 0.2  &  13.20 $\pm$ 0.02 & 12.62 $\pm$ 0.09   &   ...    &   ...       &  12.05 $\pm$ 0.03   &      ...         &       ...        &      ...        &         ...        \\
2.05714  &  4.5 $\pm$ 0.2  &  13.28 $\pm$ 0.02 & 12.72 $\pm$ 0.08   &   ...    &   ...       &  11.74 $\pm$ 0.05   &      ...         &       ...        &       ...        &        ...        \\
2.05743  &  3.9 $\pm$ 0.3  &  13.03 $\pm$ 0.02 & 12.74 $\pm$ 0.07   &   ...    &   ...       &  10.85 $\pm$ 0.37   &      ...         &       ...        &       ...        &        ...        \\
2.05800  &  6.2 $\pm$ 1.5  &  12.46 $\pm$ 0.08 &  ...               &   ...    &   ...       &  10.81 $\pm$ 0.44   &      ...         &       ...        &       ...        &        ...        \\
\hline\hline								     											      							      
2.05836  &  8.5 $\pm$ 1.3  &  12.71 $\pm$ 0.05 & 12.14 $\pm$ 0.32   &   ...    &   ...       &  11.39 $\pm$ 0.14   &      ...         &       ...        &       ...        &        ...        \\
2.05880  &  7.3 $\pm$ 0.8  &  12.90 $\pm$ 0.04 & 12.75 $\pm$ 0.08   &   ...    &   ...       &  11.65 $\pm$ 0.07   &      ...         &       ...        &      ...        &         ...        \\
2.05896  &  4.0 $\pm$ 0.2  &  13.35 $\pm$ 0.02 & 12.74 $\pm$ 0.06   &   ...    &   ...       &  11.89 $\pm$ 0.04   &      ...         &       ...        &       ...        &        ...        \\
2.05919  &  5.2 $\pm$ 0.2  &  13.57 $\pm$ 0.03 & 12.74 $\pm$ 0.05   &   ...    &   ...       &  12.13 $\pm$ 0.03   & 13.72 $\pm$ 0.06 & 13.38 $\pm$ 0.04 &       ...        & 12.55 $\pm$ 0.07  \\
2.05930  &  4.8 $\pm$ 0.1  &  14.18 $\pm$ 0.06 & 13.18 $\pm$ 0.07    &   ...    &   ...       &  11.86 $\pm$ 0.07   & 14.16 $\pm$ 0.04 & 14.34 $\pm$ 0.02 &       ...        & 13.35 $\pm$ 0.03  \\
2.05943  & 11.3 $\pm$ 0.9  &  14.11 $\pm$ 0.06 & 13.43 $\pm$ 0.07   &   ...    &   ...       &  12.73 $\pm$ 0.02   & 14.07 $\pm$ 0.07 &       ...        &       ...        & 12.78 $\pm$ 0.16  \\
2.05955  &  6.4 $\pm$ 0.2  &  13.82 $\pm$ 0.05 & 13.24 $\pm$ 0.05   &   ...    &   ...       &  12.06 $\pm$ 0.04   & 14.20 $\pm$ 0.03 & 14.00 $\pm$ 0.02 &       ...        & 13.19 $\pm$ 0.03  \\
2.05979  & 10.1 $\pm$ 0.5  &  13.45 $\pm$ 0.02 & 13.40 $\pm$ 0.04   &   ...    &   ...       &  11.79 $\pm$ 0.06   &      ...         &       ...        &       ...        & 12.62 $\pm$ 0.06  \\
2.06017  &  5.0 $\pm$ 0.2  &  13.68 $\pm$ 0.02 & 13.49 $\pm$ 0.02   &   ...    &   ...       &  11.92 $\pm$ 0.04   & 13.43 $\pm$ 0.10 &       ...        &         ...     & ...   \\
2.06047  &  6.5 $\pm$ 2.0  &  12.40 $\pm$ 0.11 &          ...       &   ...    &   ...       &      ...         &          ...        &      ...        &     ...         &         ...        \\
\hline
Total    &  & 14.69$\pm$0.02  &   14.12$\pm$0.02 & $\le$ 11.66 & $\ge$ 13.51 & 13.06 $\pm$ 0.04   &  14.70$\pm$0.02  &  14.53$\pm$0.02  &  $\ge$ 15.2 &  13.71$\pm$0.01  \\ 
\hline \hline
\end{tabular}
\caption{\label{tab:ions_single} Metal ions column densities from
Voigt profile fits.  The upper section (separated by double horizontal lines) shows fits for the
satellite complex, the lower section for the main complex, as defined
in the text. Totals in the bottom row are only for the main complex (lower section)}
\end{center}
\end{table}
\end{landscape}

\begin{table}
\centering
\caption{CIV and SiIV column densities from Voigt profile fits.  The
upper table section shows fits for the satellite complex, the lower
section for the main complex, as defined in the text.  The totals
in the bottom row also refer only to the main complex.}
\begin{tabular}{cccc}
\hline \hline
z             &  b             & log N(CIV)        & log N(SiIV)       \\
              &  (km~s$^{-1}$)      &  (cm$^{-2}$)           &  (cm$^{-2}$)           \\
\hline \hline
2.05685  &    9.0 $\pm$  0.5 &  13.32 $\pm$ 0.08 &  12.70 $\pm$ 0.07 \\
2.05695  &    6.6 $\pm$  3.2 &  13.21 $\pm$ 0.73 &  12.07 $\pm$ 1.05 \\
2.05698  &    4.1 $\pm$  0.4 &  13.18 $\pm$ 0.45 &  13.00 $\pm$ 0.09 \\
2.05704  &    4.0 $\pm$  1.2 &  13.15 $\pm$ 0.32 &  12.47 $\pm$ 0.30 \\
2.05711  &    4.7 $\pm$  0.9 &  13.05 $\pm$ 0.15 &  12.68 $\pm$ 0.11 \\
2.05721  &    7.5 $\pm$  0.9 &  13.04 $\pm$ 0.06 &  12.34 $\pm$ 0.07 \\
2.05749  &   19.5 $\pm$  1.6 &  13.12 $\pm$ 0.03 &  12.39 $\pm$ 0.04 \\
2.05800  &   18.7 $\pm$  1.0 &  13.21 $\pm$ 0.02 &  12.40 $\pm$ 0.03 \\
\hline\hline
2.05838  &   13.5 $\pm$  2.2 &  12.60 $\pm$ 0.11 &  12.01 $\pm$ 0.10 \\
2.05879  &    6.3 $\pm$  0.4 &  12.75 $\pm$ 0.03 &  12.22 $\pm$ 0.04 \\
2.05916  &   55.4 $\pm$ 11.8 &  13.48 $\pm$ 0.17 &  12.68 $\pm$ 0.14 \\
2.05923  &    6.6 $\pm$  0.2 &  13.83 $\pm$ 0.05 &  13.30 $\pm$ 0.04 \\
2.05930  &    8.8 $\pm$  1.2 &  13.96 $\pm$ 0.18 &  13.38 $\pm$ 0.23 \\
2.05943  &   28.6 $\pm$  7.4 &  13.50 $\pm$ 0.10 &  12.88 $\pm$ 0.13  \\
2.05947  &   12.8 $\pm$  3.8 &  13.68 $\pm$ 0.34 &  13.23 $\pm$ 0.32 \\
2.05978  &   12.1 $\pm$  1.9 &  12.95 $\pm$ 0.16 &  12.37 $\pm$ 0.12 \\
2.06018  &    6.3 $\pm$  1.2 &  11.78 $\pm$ 0.29 &  11.83 $\pm$ 0.07 \\ 
2.06033  &   28.4 $\pm$  3.2 &  13.08 $\pm$ 0.07 &  11.96 $\pm$ 0.13 \\ 
\hline   
\multicolumn{2}{l}{Total}      &  14.48 $\pm$ 0.01 &  13.86 $\pm$ 0.01 \\ 
\hline \hline
\label{tab:ions_high}
\end{tabular}
\end{table}

The same kinematic structure (i.e. combination of Doppler ($b$)
parameter and $z$) was used for singly ionized species, \ion{N}{i} and
\ion{Al}{iii}, but this model was not applicable to the more highly
ionized species of \ion{C}{iv} and \ion{Si}{iv}.  We also adopted an
independent model for \ion{S}{i} and \ion{C}{i}, since these atoms
likely trace gas in a cooler kinematic phase.  The fits for the
measured species are given in Tables~\ref{tab:ions_neutral} to
\ref{tab:ions_high} and shown in Figures \ref{fig:CIfig} and
\ref{fig:metfig}.

\subsection{Molecular hydrogen}

Molecular transitions of hydrogen from both Lyman and Werner bands up to
rotational level $J=5$ are detected in two kinematically distinct components, at
the velocity of the strongest metal complex. The H$_2$ lines are located at the
same redshifts as the neutral lines of carbon, which are also aligned with the
strongest singly ionized metal ion lines.  Fitting the H$_2$ absorption
features is complicated by saturation of the lowest J states and by
the need for continuum and zero-level adjustment.
Molecular HD is also detected in this system, only the third such detection in a
DLA (see also Varshalovich et al 2001; Srianand et al. 2008).  The HD and H$_2$
lines are analysed in two separate papers (Malec et al. 2010; Tumlinson et al.
in preparation).  The molecular lines are not considered further in this paper,
but their presence is noted here for completeness.

\section{Chemical abundances, molecular fraction and cooling rate}\label{abund_sec}

\begin{table}
\begin{center}
\caption{Elemental abundances before ([X/H]$_{raw}$) and after
([X/H]$_{corr}$) ionization corrections (IC(X/H).  Values are for the
main complex whose \nhi=19.18$\pm$0.15 \cm. Errors in [X/H] account
for the error in \nhi\ and N(X) and are added in quadrature.  For
presentation purposes, the errors are only quoted for the final corrected
abundance, but the errors on the raw abundance are identical. }
\begin{tabular}{lcccc}
\hline \hline
   & ${\log}$~N(SiII)  & ${\log}$~N(FeII)    & ${\log}$~N(SII)  & ${\log}$~N(NI)  \\
  &  (cm$^{-2}$) &  (cm$^{-2}$)  &  (cm$^{-2}$)   &  (cm$^{-2}$) \\
\hline
Total  & 14.69$\pm$0.02 &  14.12$\pm$0.02    &  14.70$\pm$0.02  &  14.53$\pm$0.02     \\ 
\hline \hline
$\log ({\rm X}/{\rm H})_\odot$  & $-4.49$  &    $-4.55$  &  $-4.84$  &  $-4.22$ \\ 
$[{\rm X}/{\rm H}]_{raw}$  &  $+0.00$  & $-0.51$   & $ +0.36$	&  $-0.43$ \\ 
IC(${\rm X}/{\rm H}$) & $+0.71$ & $+0.26$    & $+0.55$  &  $-0.04$  \\ 
\hline
$[{\rm X}/{\rm H}]_{corr}$ & $-0.71\pm0.15$ &  $-0.77\pm0.15$    & $-0.19\pm0.15$  &  $-0.39\pm0.15$ \\ 
\hline \hline
\label{tab:abund}
\end{tabular}
\end{center}
\end{table}

The abundances, calculated on the basis of the raw column density
measurements (uncorrected for ionization), in the sub-DLA towards
J2123$-$0050 reveal a number of surprising results.  These values,
calculated relative to the solar abundance pattern of Grevesse,
Asplund, \& Sauval (2007), are given in Table \ref{tab:abund}. 
We consider only the main absorption complex in this analysis, using an
\nhi\ = 19.18.

\subsection{Metallicity Without Ionization Corrections}

Uncorrected for ionization, the metallicity of the sub-DLA (main
complex) appears to
be super-solar, i.e. [Si/H] = $+0.00$ and [S/H] = $+0.36$. Super-solar
metallicities have previously been presented for a small number of
sub-DLAs in the literature (e.g. Prochaska et al. 2006; Peroux et
al. 2006; Peroux et al.  2008; Meiring et al. 2008; Dessauges-Zavadsky
et al. 2009), based on a similar analysis (i.e. assuming small or
negligible ionization corrections).  Such high metallicities are
somewhat surprising at high redshift for a number of reasons.  First,
the emission line abundances measured from actively star-forming
galaxies at z $>$ 2 are almost exclusively sub-solar (e.g. Erb et
al. 2006; Maiolino et al. 2008).  It is feasible that absorption and
emission line abundances yield discrepant results (e.g. Ellison,
Kewley \& Mallen-Ornelas 2005) due to abundance gradients and/or local
enrichment.  However, although the magnitude of \ion{H}{ii}-to-\hi\
region abundance disagreements is still debated (e.g.  Lebouteiller et
al. 2009 and references therein), it would be expected that any
discrepancy would tend towards higher abundances for emission
lines. Moreover, emission line abundances measure only the
metallicities of the actively star-forming regions, whereas absorption
line measurements probe the entire galaxy along a given sightline and
therefore yield average ISM measurements.  Even in cases where a
sightline probes regions near active star-formation, such as the case
for GRB-absorbers, metallicities are typically around 1/10 Z$_{\odot}$
(Prochaska et al. 2007a).

\subsection{Relative Abundances Without Ionization Corrections}

Although the range of elements probed by our spectra is somewhat
limited, there are two notable abundance ratios of interest.  The
first are indicators of $\alpha$ element enhancement, which can be
obtained from the ratios of [S/Fe]= +0.87 and [Si/Fe] = +0.51.
The complexities of interpreting such ratios in the context of nucleosynthetic
versus dust effects have been discussed by many authors (e.g. the
review by Wolfe et al. 2005).  What is striking in the case of J2123$-$0050
is the extremely high values of [S,Si/Fe] compared to previous measurements
(e.g. Prochaska \& Wolfe 2002).  The main complication here is the
unquantified effect of dust depletion, particularly for \fe2\ which can
raise these ratios far above their intrinsic values.  

A second ratio which reveals a surprising result is N/S.  The utility
of nitrogen as a cosmic clock, thanks to its primary and secondary
contributions, has been discussed extensively in the literature
(e.g. Pettini et al. 2002, Prochaska et al. 2002a; Centurion et
al. 2003; Henry \& Prochaska 2007; Petitjean et al. 2008).  In brief,
the dominant contribution of $\alpha$ elements to the ISM comes from
prompt enrichment by massive stars which result in Type II SN.
Secondary nitrogen is released on a longer timescale, once the seed
nuclei of carbon have been established from previous generations of
stars.  Hence, the secondary component of nitrogen steadily builds
with time once the metallicity is approaching the solar value.  The
source of primary nitrogen is still a contentious issue, but the
substantial scatter of N/$\alpha$ in DLAs has led to the suggestion
that primary nitrogen production occurs in low or intermediate mass
stars and its release is hence is delayed relative to the $\alpha$
elements (Pettini et al. 2002; Henry \& Prochaska 2007).

\begin{figure}
\centerline{\rotatebox{0}{\resizebox{8cm}{!}
{\includegraphics{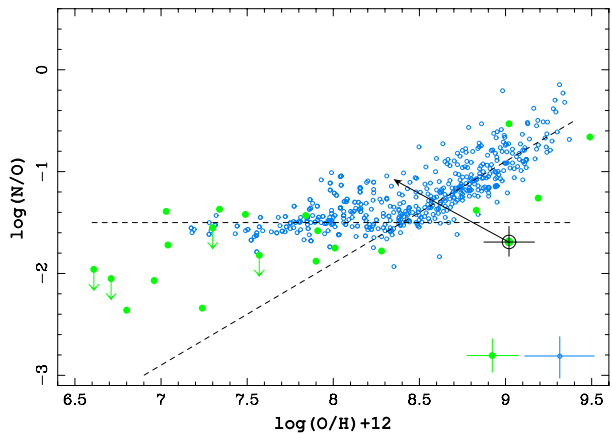}}}} 
\caption{\label{fig:ns} N/O ratio plotted against the
system metallicity on the solar scale.  Metallicities are determined
from O/H or Si, S/H corrected by the solar ratio of Si/O or S/O (see
Table \ref{tab:abund}).  The sub-DLA towards J2123$-$0050 (ringed
green circle) is plotted against the DLA data (green circles) and
Galactic \ion{H}{ii} regions (open blue circles) assembled by Pettini
et al.  (2008). The dashed lines show predicted contributions from
primary and secondary N production.  The arrow points to the
J2123$-$0050 ratio corrected for ionization. Representative error
bars for HII regions and DLAs are given in the lower right of the figure.}
\end{figure}

The sub-DLA towards J2123$-$0050 is the first QSO absorber with solar
or super-solar metallicity in which N has been measured.  At such high
metallicities, Galactic \ion{H}{ii} regions show N/$\alpha$ ratios
that are correlated with metallicity (usually measured from O/H) as
expected from the secondary production mechanism described above.  In
Figure \ref{fig:ns} we show the ratio of N/$\alpha$ in the
J2123$-$0050 sub-DLA along with other DLA N measurements from the
literature and Galactic \ion{H}{ii} regions.  The sub-DLA lies in a
previously unpopulated part of the diagram -- high metallicity, but
relatively low N/$\alpha$.  This is a puzzling result.  On the one
hand, this DLA has apparently experienced sufficient star formation to
enrich its ISM to above solar values, but is only experiencing primary
nitrogen enrichment.  One explanation of this combination of
abundances would be that the galaxy is both chemically young, but has
experienced intense and productive star formation.

\subsection{Cooling rate}

The sub-DLA towards J2123$-$0050 is also an outlier in the cooling
rate distribution of DLAs.  Wolfe et al. (2003) derived a method to
determine the properties of the star forming regions associated with
the DLA sightlines through the detection of the
\ion{C}{ii}$^{*}~\lambda$1335 line. The method estimates
[CII]~158$\mu$m emission from the strength of the
\ion{C}{ii}$^{*}~\lambda$1335 absorption. Absorption resulting in the
1335~{\AA} line arises from $^2$P$_{3/2}$, from which a 158 $\mu$m
photon is spontaneously emitted during the decay to the $^2$P$_{1/2}$
fine-structure state in the ground 2s$^{2}$2p term of \ion{C}{ii}. The
[\ion{C}{ii}]~158$\mu$m transition is a principal coolant of
interstellar neutral gas in the Galaxy (Wright et al.  1991). By
assuming that cooling is dominated by the 158$\mu$m line, the heating
rate can be calculated by determining the [\ion{C}{ii}] cooling rate
$l_c$.

The emissivity of [\ion{C}{ii}]~158$\mu$m can be calculated from the
column density of \ion{C}{ii}$^{*}~\lambda$1335 line following the
expression of Pottasch, Wesselius, \& van Duinen (1979):

\begin{equation} l_{c} = \frac{N(CII^{*}) h\nu_{ul}A_{ul}}{N(HI)}~{\rm ergs s^{-1}
Hz^{-1}} \label{eq:lc} 
\end{equation} 

\noindent where $A_{ul}$ is the Einstein A coefficient, and $h\nu_{ul}$ is the
energy of the 158$\mu$m transition. For the sub-DLA towards
J2123$-$0050, we derive $l_c = 1.03\times~10^{-25}$~
ergs~s$^{-1}$~Hz$^{-1}$.  Wolfe et al. (2008) have recently shown that
$l_c$ in DLAs follows a bimodal distribution with a transition at $l_c
\sim 10^{-27}$~ ergs~s$^{-1}$~Hz$^{-1}$.  Absorbers above this
critical value exhibit higher metallicities, velocity widths and
dust-to-gas ratios.  Furthermore, Wolfe et al. (2008) suggest that
galaxies with cooling rates higher than the critical value are located
in more massive halos with active star formation occuring in `bulge
mode', i.e. removed from the gas halo.  Although the sub-DLAs towards
J2123$-$0050 qualitatively follows the above trend with its
super-solar metallicity and complex velocity structure, it has a
cooling rate that exceeds all the values in the Wolfe et al. (2008)
DLA sample by a factor of ten (although saturation of the \ion{C}{ii}
leads to lower limits for $l_c$, especially at high \nhi), and is also
larger than the average Galactic disk value.

\section{Evidence for a multi-phase medium}\label{mp_sec}

The neutral inter-stellar medium is generally divided into two
components -- a warm neutral medium (WNM) with temperatures of several
thousand degrees K and densities $\sim$ 0.1 atoms cm$^{-3}$, and the
cold neutral medium (CNM) with T $\sim$ 100 K and $n \sim 10$ atoms
cm$^{-3}$ (Field et al. 1969).  Spin temperatures of high redshift
DLAs are generally high (e.g.  Kanekar et al. 2006 and references
therein), with only one DLA exhibiting a value below 350 K at $z_{\rm
abs} > 1$ (York et al. 2007).  These high values are consistent with
sightlines that intersect both CNM and WNM gas.  Kanekar, Ghosh \&
Chengalur (2001) show this temperature segregation explicitly for one
DLA at $z_{\rm abs} \sim 0.2$.  At higher redshift, absorption from a
range of high and low ionization species indicates that a multi-phase
structure in DLAs is common (e.g. Wolfe \& Prochacka 2000; Fox et
al. 2007; Lehner et al. 2008; Quast et al.  2008).  In this section,
we examine the evidence for a multi-phase medium in the sub-DLA
towards J2123$-$0050.

The sub-DLA towards J2123$-$0050 is a kinematically complex system
with metal absorption spreading over 400 \kms\ in velocity space
requiring 15 components for a reasonable Voigt profile fit, see Figure
\ref{fig:metfig}.  Although such a wide velocity spread is not unique
amongst quasar absorbers, this sub-DLA is unusual in the range of
species that are detected.  In addition to the often-observed singly
ionized species, such as \ion{Si}{ii} and \ion{Fe}{ii}, the much less
common \ion{C}{i} line is present and a rare detection of \ion{S}{i}
(see also Quast et al. 2008).  Substantial column densities of more
highly ionized species are also measured, for example, for
\ion{Al}{iii}, \ion{C}{iv} and \ion{Si}{iv}.  The singly and multiply
ionized metal species are clearly separated into a main complex
centred at $z_{\rm abs} = 2.05934$ and a satellite complex at $z_{\rm
abs} = 2.05684$ (i.e. separated by $\sim$ 250 \kms).  The neutral
species, such as \ion{C}{i} and \ion{S}{i} are detected in only two
components of the main complex.  The same is true for the molecular
absorption (Malec et al. 2010; Tumlinson et al. in prep) which is
observed in only one (HD) or two (H$_2$) velocity components
associated with the main complex.  The simultaneous presence of such a
variety of ionized species for a given element and the kinematic
diversity is strongly suggestive of a multi-phase medium in which
neutral and molecular species occupy only a fraction of the
interstellar volume.

Further evidence for a multi-phase medium comes from the details of
the Voigt profile fit.  The neutral carbon and sulphur lines require a
very small $b$-parameter of only 1.8 \kms.  These very narrow lines
are unresolved even in this high resolution spectra
($R\sim100,000$). However, because of a high range of $gf$ values in
which \ion{C}{i} transitions are observed, the b-parameters of these
lines are constrained well. The singly ionized species are more than
twice as broad with $b$-parameters $\sim$ 4.8 \kms, such differences
are common signatures in the Galactic ISM (e.g. Spitzer \& Jenkins
1975).  Ignoring the contribution of turbulence in line broadening,
for a given chemical element that is detected in multiple ionization
stages, the ratio of Doppler parameters scales as the square root of
the ratio of temperatures, a factor of seven in the case of neutral
and singly ionized sulphur.  The sub-DLA studied here therefore
presents a clear case of a multi-phase medium at high redshift.
 
\section{Ionization models for a two-phase medium}\label{cloudy_sec}

As discussed in the Introduction, ionization corrections are usually
assumed to be negligible for DLAs, and have also been reported to be
$<0.2$ dex in sub-DLAs.  Although this value is not large, corrections
of this magnitude may be sufficient to mask nucleosynthetic signatures
(e.g. Prochaska et al. 2002a; Quast et al. 2008).  Previous attempts
to model ionization corrections have assumed a single phase model. Our
observations of J2123$-$0050 clearly demonstrate that this is not
always an accurate assumption.  However, single phase models are
useful because they do not require knowledge of how much of the
observed \nhi\ is associated with each phase.  As discussed by Ellison
et al. (2007), it is extremely difficult to separate DLA \lya\
profiles into separate components\footnote{At lower column densities,
asymmetries in the \lya\ wing become more obvious and separation
becomes easier}, although species such as O~I may provide clues in
this respect (e.g. Fox et al. 2007).  To assess the impact of the
single-phase assumption we use Cloudy (Ferland et al. 1998) to
construct a multi-phase model and then test how accurately the column
densities can be recovered under the single-phase assumption.

The modelling procedure is as follows.

\begin{enumerate}

\item A model with two phases is constructed, where the `phase' is
defined by the ionization parameter, i.e., the ratio of H ionizing
photons to H atoms.  The `cold' phase gas is defined as having an
ionization parameter of ${\log}U=-5.0$, while the `warm' phase is set
to ${\log}U=-3.0$. The total combined neutral column density of the
cloud in both phases is set to be \nhi\ =$10^{19.18}$~\cm.  Nine
versions of the model were constructed with different fractions of the
\nhi\ in the warm and cold phases ranging from completely cold to
completely warm gas (essentially these extrema are single-phase
clouds). 

\item The metallicity of the cloud is fixed at [M/H]=$-0.33$ in all
models, with a solar abundance pattern as given in Cloudy version
07.02.

\item The two-phase cloud is radiated with a mix of the Haardt-Madau
(H\&M) extragalactic spectrum (Haardt \& Madau 1996) and the average
Galactic ISM spectrum of Black (1987), which primarily affects the
ions with ionization potential lower than 1Ryd, such as \ion{C}{i}, and
\ion{S}{i}.  

\item  For each of the nine models (which sample different fractions of
cold and warm gas),  Cloudy outputs the column
densities of \fe2, \s2, \si2, \ion{Al}{ii} and \ion{Al}{iii} separately
for the cold and warm phase.  The column densities of a given
species in the cold and warm phases are summed to give the
column densities that would be observed in a real spectrum.

\end{enumerate}

The procedure described above yields nine sets of column densities for
two-phase models.  The next step is to recover the metallicity of the
theoretical input cloud by following the steps that an observer would
execute in trying to model the cloud with a single phase.  This
requires adopting a metallicity and an indicator of log U and then
running Cloudy over a parameter grid until the observed values are
reproduced.  All grids are calculated using a mix of the
H\&M and the average Galactic ISM spectrum stopping the calculations
when the column density of neutral hydrogen reaches \nhi\
=$10^{19.18}$~\cm.  The details of this stage in the modelling are:

\begin{enumerate}

\item  For the metallicity, we assume [M/H]=[\fe2/\hi] (where
the column densities come from the Cloudy output).  In order to
test whether this approximation of metallicity introduces a significant
error, we also repeat the experiment with the `true' metallicity
of [M/H]$=-0.33$ (in practice, this is unknown to the observer).
	
\begin{figure}
\centerline{\rotatebox{0}{\resizebox{10cm}{!}{\includegraphics{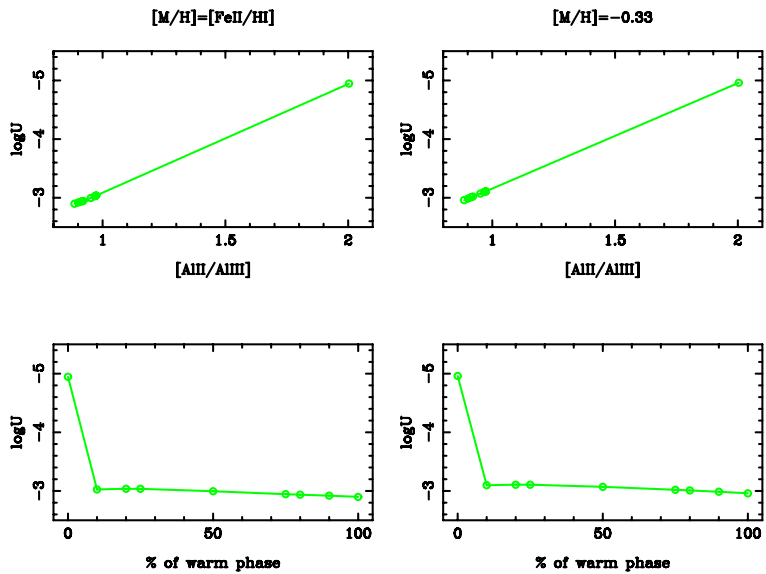}}}}
\caption{Derivation of the ionization parameter from aluminium ionic
ratios calculated for theoretical two phase clouds, assuming a single
phase. The top panels show a monotonic relation between
\ion{Al}{ii}/\ion{Al}{iii} and the inferred ${\log}~U$.  The lower
panels show the inferred ionization parameter as a function of the
fraction of N(\hi) in the warm phase. The left hand column is for the
single-phase models that assume [M/H]$=-0.33$, while the right is for
models with [M/H]=[\fe2/\hi].  Even if the contribution of the warm
phase to the total column density of \hi\ is as small as 10\% the
models recover the $\log~U$ of the warm phase to within
$\sim$0.10~dex.\label{fig:Al_models}}
\end{figure}
 
\item The ionization parameter of the single-phase model of each of
the theoretical clouds is inferred from the column density ratio of
\ion{Al}{ii}/\ion{Al}{iii}.  This ratio is commonly used as an
indicator of ionization in sub-DLAs and DLAs (Dessauges-Zavadsky et
al. 2003; Prochaska et al 2002b).  One of the empirical motivations
for using \ion{Al}{ii}/\ion{Al}{iii} is its correlation with neutral
hydrogen column density (Vladilo et al 2001, Dessauges-Zavadsky et al
2002). Narayanan et al (2008) speculated that this anti-correlation
might extend also to the lower column denisty QALs, such as weak
\ion{Mg}{ii} absorbers.  However, other work has questioned the
utility of \ion{Al}{ii}/\ion{Al}{iii} for constraining the ionization
parameter (e.g. Howk \& Sembach 1999), whose atomic data is
additionally poorly known (e.g. Vladilo et al. 2001;
Dessauges-Zavadsky 2003).  However, since our methodology is to adopt
standard observational strategies to quantify errors in derived
properties, we follow the common practice of using
\ion{Al}{ii}/\ion{Al}{iii} to constrain ionization parameter.
Figure~\ref{fig:Al_models} presents the inferred value of log $U$ for
each of the nine theoretical two-phase clouds.  The left-hand panels show
the results of the models in which the metallicity is estimated from
\fe2/\hi, whereas the right-hand panels show the models using the
correct metallicity.  The left-hand panels are almost identical to
those on the right, indicating that, for these models, using \fe2/\hi\
as an estimate for metallicity is a good approximation. As expected,
\ion{Al}{ii}/\ion{Al}{iii} increases monotonically with
${\log}~U$. Perhaps more surprising is that this approach fairly
faithfully recovers the ionization parameter of the warm phase even if
its contribution to the total neutral hydrogen density is as small as
10\%. The difference between the real and recovered ${\log}~U$ for the
warm phase is only on the order of 0.10~dex.
	
\begin{figure}
\centerline{\rotatebox{0}{\resizebox{8cm}{!}{\includegraphics{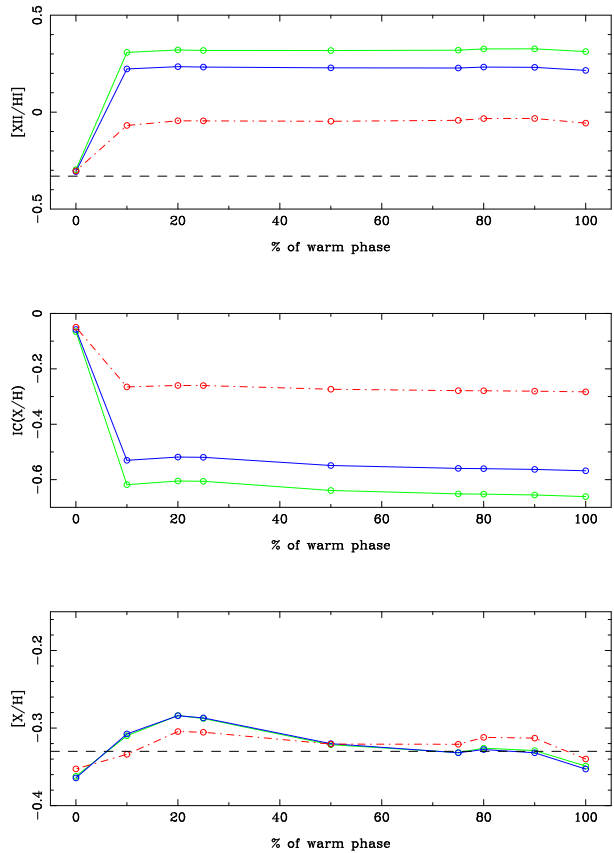}}}}
\caption{\label{fig:ex1} Ionization corrections for the theoretically
modeled medium with [M/H]=$-0.33$. The ionic abundances `measured'
from the Cloudy models are presented in the top panel for \si2\ (green
line), \s2\ (blue line) and \fe2\ (red dotted line). The ionization
corrections are given in the middle panel, and the corrected
abundances are shown in the bottom panel. The black horizontal line
represents the input metallicity of [M/H]$=-0.33$. The single-phase
based corrections recover the metallicity of the clouds to within
$\sim$0.15~dex.}
\end{figure}

\begin{figure}
\centerline{\rotatebox{0}{\resizebox{8cm}{!}{\includegraphics{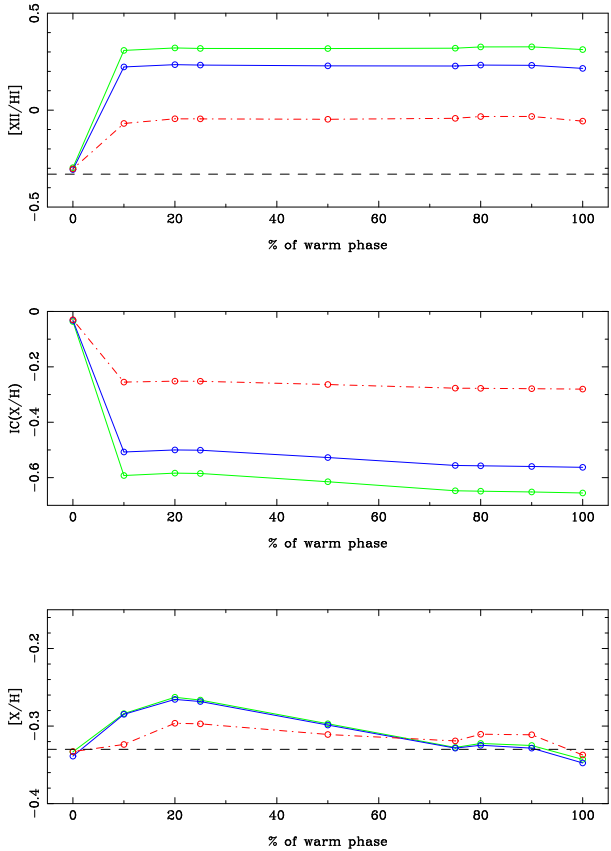}}}}
\caption{\label{fig:ex2} Same as in Figure~\ref{fig:ex1}, but for an
assumed metallicity of [M/H]=[\ion{Fe}{ii}/\ion{H}{i}].}
 \end{figure}
	
\item After the ionization parameter is derived for each model, the
ionization corrections are calculated. The fractional column density
of an element in a given ionization state is given as $f({\rm
X}^{i+})=\frac{N({\rm X}^{i+})}{N({\rm X})}$, and similarly for
hydrogen $f(HI)=\frac{n({\rm HI})}{n({\rm H})}=\frac{N({\rm
HI})}{N({\rm H})}$. The ionization corrections $IC$(X/H) are then
given by:

\begin{equation}
IC({\rm X}/{\rm H}) = \log\left(\frac{N({\rm X}^{i+})}{N({\rm HI})}\right) - \log\left(\frac{N({\rm X})}{N({\rm H})}\right),
\end{equation}

\noindent which is simply:

\begin{equation}
\label{IC_eqn}
IC({\rm X}/{\rm H}) = \log\left(\frac{f({\rm X}^{i+})}{f({\rm HI})}\right),
\end{equation}

\noindent These values are subtracted from the ionic abundances to
obtain the final abundance of an element.  Comparing these corrected
abundances to the input values of the original two-phase model allows
us to assess how accurately ionization corrections for a single phase
model recover the input abundances of a truly two-phase medium.

\end{enumerate}

The results of this experiment are presented in Figure~\ref{fig:ex1}
for the case where the intrinsic metallicity is known, and
Figure~\ref{fig:ex2}, for the case which uses the `observed'
metallicity. The top panel presents the `observed' ionic abundance for
\si2, \s2 and \fe2, the middle panel shows the ionization corrections,
and the bottom panel gives the corrected metallicities.  The black
line indicates the input metallicity of [M/H]$=-0.33$.  Only the
results for \si2, \s2 and \fe2\ are shown, as these are the ions
meausred in our study.  For completeness, we also show in the Appendix
a more complete set of models for a further six elements commonly
measured in DLAs.  The recovered metallicity for all of the two-phase
models over-estimates the input value.  However, the deviation from
the [M/H]$=-0.33$ value is less than 0.15~dex for both metallicity
models, with the error for Fe consistently lower than that of Si, and
S. The accuracy of the corrections increases with increasing
warm-phase fraction. The input metallicity is recovered when the model
is either 100\% warm or 100\% cold phase, since in these cases, the
single-phase approximation is obviously an accurate one.  From these
models, it is possible to quantify the approximate accuracy to within
which ionization corrections can be calculated from a single phase
Cloudy model. The experiment also shows that if even a small fraction
of hydrogen column density is present in the warm phase, applying no
corrections to the elemental abundances can lead to over-estimates up
to 0.5~dex (as is the case for Si in this example).

\section{Corrected column densities for J2123$-$0050}\label{sec_corrections}

In this section we derive photo-ionization corrections for the sub-DLA
towards J2123$-$0050 based on a single-phase Cloudy model and present
the corrected elemental abundances.

\subsection{Cloudy model for J2123$-$0050}

As shown in Section \ref{cloudy_sec}, for a multi-phase medium, the
corrections derived from the single-phase Cloudy model approximate the
values of the warm medium and become increasingly inaccurate as the
medium becomes dominated by a cooler phase.  It is unfortunately not
possible to model the corrections for J2123$-$0050 with a multi-phase
medium since the distribution of neutral hydrogen between the phases
is impossible to constrain.  For the case simulated above, we have
shown that although ionization corrections that assume a single phase
medium may still tend to over-estimate the intrinsic abundances, this
effect is typically at the $<$0.1 dex level.  Applying no ionization
correction can lead to abundance over-estimates of more than 0.5 dex.
We therefore take the enforced approach of adopting a single-phase
model, but with the knowledge that further corrections for the
multi-phase structure are likely to be small.  Additional sources of
uncertainty present in the modelling of any given absorber with Cloudy
include assumptions for the UV background, relative abundances, atomic
data and absorber geometry.

\begin{figure}
\centerline{\rotatebox{0}{\resizebox{8cm}{!}
{\includegraphics{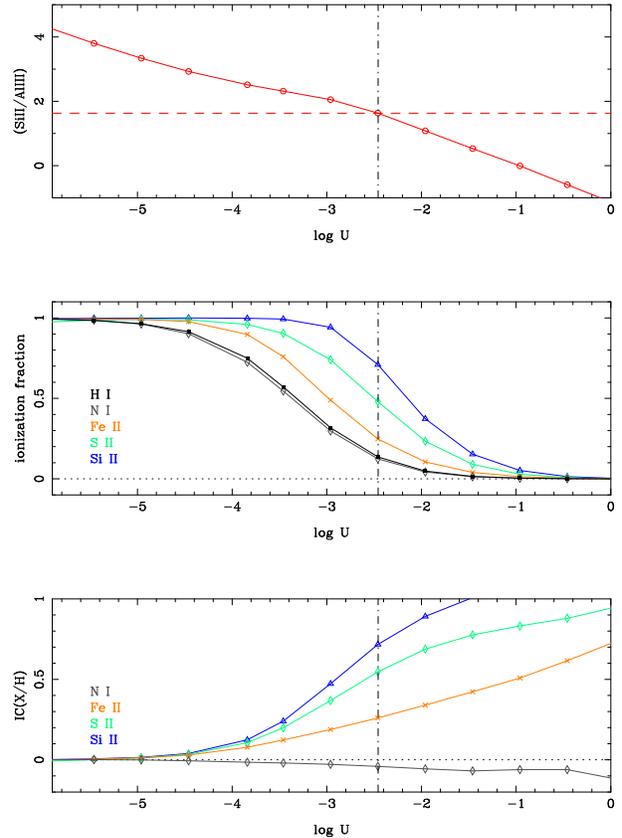}}}}
\caption{\label{fig:ion_corr} The top panel shows the
predicted \si2\ to \ion{Al}{iii} column density ratio from the Cloudy models
with the photoionizing spectrum containing both average Galactic ISM and
extragalactic HM spectra. The horizontal dashed lines is drawn at the
measured ratio.  The average ionization parameter inferred by comparing the
observed values to the model has a value of $-2.46$.  The middle
panel shows the fraction of each element in the given stage of ionization.
Predicted ionization
corrections for the selected metal abundances (see equation \ref{IC_eqn}) 
from the same models are given in the lower panel.  }
\end{figure}

To calculate the ionization corrections (eqn \ref{IC_eqn})
for the sub-DLA J2123$-$0050
we constrain the Cloudy model to have \nhi\ = $10^{19.18}~$\cm, a
metallicity of +0.36 (from the raw measurement of \ion{S}{ii} and
\hi), and a solar abundance pattern.  In order to determine the
ionization parameter (which, as we have shown above, approximates to
that of the warm medium if multiple phases are present), the code
matches the observed (\si2/\ion{Al}{iii}) ratio of 1.63 to the values
obtained from the model (see top panel of Figure~\ref{fig:ion_corr}).
\si2\ is used as a proxy for \ion{Al}{ii} since the only available
\ion{Al}{ii} line is saturated and only a lower limit to aluminium
ionic ratio (\ion{Al}{ii}/\ion{Al}{iii}$>0.45$) can be measured from
the available spectrum.  

The upper panel of Figure~\ref{fig:ion_corr} presents the ionic ratio
(\si2/\ion{Al}{iii}) versus the ionization parameter from the model
grid. The observed ratio of ions intercepts the model curve at the
ionization parameter of ${\log}U = -2.46$. The lower panels give the
model inferred ionization fraction and correction curves for the ions
of interest.  Table \ref{tab:abund} reports the ionization corrections
(IC) for each element, and the corrected abundance on the solar scale
once the ionization has been accounted for.  The errors on the column
densities are those determined by VPFIT (Table \ref{tab:ions_single}),
and errors on abundances include the error in \nhi\ propogated in
quadrature.  No attempt has been made to estimate the error on the
ionization correction due to the uncertainties mentioned above.
Nitrogen (relative to \hi) is the least
affected by ionization, which is often implicitly assumed, due to its
ionization potential of 1.07 Ryd.  Indeed, the middle panel of Figure
\ref{fig:ion_corr} shows that the fraction of \ion{N}{i} closely
tracks the \hi\ as a function of log U.  \si2\ is the species
requiring the largest ionization correction for the limited species
studied here (see the Appendix for corrections for other elements).

In order to examine the influence of the photoionizing spectrum and
the assumed metallicity on the magnitude of the inferred corrections,
additional Cloudy grids of varying metallicity (1/3 solar, and solar)
and photoionizing radiation field (by scaling the Galactic ISM spectra
component to 0, 1/10, 1/3, 3, and 10 of the original) were
produced. The models with the combination of these parameters that
acceptably reproduce the observed elemental abundance ratios vary the
magnitude of the resulting correction within a couple of tenths of a
dex from the value determined by the original model.

\subsection{Application of ionization corrections}

We now re-examine the metallicity, cooling rate, N/$\alpha$ and
$\alpha$/Fe abundances of the sub-DLA towards J2123$-$0050 using
ionization corrected metal abundances.  Although the `raw' abundances
indicated that this sub-DLA had a super-solar abundance in both Si and
S, the corrected abundances of all elements are below the solar value:
[S/H] = $-0.19$ and [Si/H] = $-0.71$.  This sub-DLA therefore remains
metal-rich for its redshift, relative to DLAs at $z \sim 2$,
superficially supporting the idea (e.g. Peroux et al. 2007) that
sub-DLAs may be preferentially more metal-rich than DLAs.  However, a
recent re-appraisal of this hypothesis by Dessauges-Zavadsky et
al. (2009) has argued that the difference in metallicity only exists
at low redshift ($z_{\rm abs} < 1.7$), and have suggested that this
might be due to a selection effect.  In the case of the sub-DLA
towards J2123$-$0050, the comparison with the bulk of DLAs is not a
fair one, since this system was selected to be metal-strong.
Herbert-Fort et al. (2006) provided the first evidence that absorbers
selected based on strong metal lines have a tendency to approach solar
metallicity.  This has been confirmed by Kaplan et al. (2010) who find
typical metallicities of metal-strong DLAs at $z_{\rm abs} \sim 2$ in
the range of 1/10 to 1/3 of solar.  The sub-DLA studied here is
therefore consistent with the range of metallicities measured in
identically selected, higher \hi\ column density systems at the same
redshift.

The determination of the cooling rate includes the \nhi\ column
density; for most Galactic interstellar sightlines the total hydrogen
content is indeed dominated by neutral gas so that \nhi\ is a
reasonable proxy in the calculations (e.g. Pottasch et
al. 1979). However, we have shown that this is not the case for the
sub-DLA studied here.  We therefore re-calculate the cooling rates as:

\begin{equation} l_{c} = \frac{N(CII^{*}) h\nu_{ul}A_{ul}}{N(H)}~ergs s^{-1}
Hz^{-1}
\end{equation} 

\noindent where the total hydrogen column density determined from our ionization
model is log N(H) = 20.05 \cm.  The cooling rate calculated for
J2123$-$0050 decreases from $l_c = 1.03\times~10^{-25}$~
ergs~s$^{-1}$~Hz$^{-1}$ to $l_c = 1.39\times~10^{-26}$~
ergs~s$^{-1}$~Hz$^{-1}$, consistent with the `high-cool' population
studied by Wolfe et al. (2008).  

Finally, the corrected N/O ratio has also moved dramatically relative
to its `raw' values, due to both corrections in metallicity and N/O
(determined from N/S with a correction for the S/O solar
abundance). The corrected value (see Figure \ref{fig:ns}) is now
indicative of a primary plus secondary nitrogen contribution and is
consistent with Galactic \ion{H}{ii} regions at a similar
metallicity. Concern regarding ionization corrections of N/$\alpha$
ratios has been previously raised, e.g. Izotov et al. (2001).
Specifically, ionization effects have been appealed to in order to
explain the large scatter of N/$\alpha$ ratios below the primary
plateau.  Although we have shown that a significant correction is
required for the sub-DLA studied here, Pettini et al.  (2002) have
argued that such corrections are unlikely to be responsible for all of
the observed scatter.

\section{Summary and discussion}\label{summary_sec}

The ISM is expected to exhibit a multi-phase structure and there is
abundant evidence that DLAs are no exception.  In this paper we
present the case of a $z_{\rm abs} = 2.06$ sub-DLA which exhibits a
complex kinematic and ionization structure.  The HIRES data obtained
represent one of the highest resolution spectra obtained of a high $z$
QSO which facilitates the study of the multi-phase ISM.  Absorption
transitions detected in this sub-DLA range from species which trace
cold gas (such as H$_2$, HD, \ion{S}{i}, \ion{C}{i}) up to highly
ionized species such as \ion{C}{iv}.  Ignoring ionization corrections
in this system leads to puzzling properties.

We have shown that using a single phase model to derive ionization
corrections in a multi-phase medium can only recover abundances to
within $\sim$ 0.15 dex (for the model example investigated
here). Although the exact accuracy of corrections will depend on a
large number of factors (input ionizing spectrum, number of phases,
their ionization parameters etc.), our experiment illustrates that
caution is required when interpreting abundances in sub-DLAs.  For
coarse properties, e.g. distinguishing a solar metallicity galaxy from
one a metal-poor one with $Z \sim 1/10 Z_{\odot}$, the single-phase
assumption probably does not impact greatly on the
conclusions. However, relative abundances, which are used to identify
subtle nucleosynthetic effects, may be quite susceptible to ionization
errors.  For example, although our two-phase model was constructed
with a solar ratio of Si/Fe, the output abundances (see Figures
\ref{fig:ex1} and \ref{fig:ex2}) yielded an over-abundance of Si by up
to 0.1 dex.  This might otherwise be interpreted as evidence for dust
depletion in Fe, or $\alpha$-element enhancement in Si.

In closing, we note that the sub-DLA studied here is of particularly
low \nhi, and is unusual in the range of ionzation species it
displays.  It may therefore not be typical of the sub-DLA population
as a whole.  However, it does alert us to the real need to assess the
ionization state of sub-DLAs on a case by case basis.  For example,
Quast et al. (2008) recently revisited the $z_{\rm abs} = 1.15$
absorber towards HE0515$-$44.  Originally estimated by de la Varga et
al.  (2000) to have log \nhi\ $\approx$ 20.45, Hubble Space Telescope
spectra yield log \nhi\ = 19.9 \cm\ (Reimers et al. 2003).  As for the
sub-DLA studied here, Quast et al. (2008) detect a range of ionzation
species including \ion{S}{i}, \ion{Si}{i}, \ion{Fe}{i}, \ion{Si}{iii}
and \ion{Al}{iii} which confirm that the sightline intersects a
combination of neutral and ionized gas.  In fact, Quast et al. (2008)
conclude that the majority of the metal line species in the sub-DLA
towards HE0515$-$44, similarly to the one in J2123$-$0050, are formed
in the ionized component.  A second interesting example is the
absorber at $z_{\rm abs} = 0.745$ towards Q1331+17.  Ellison et
al. (2003) previously investigated the complex kinematic structure and
ionization structure of this absorber.  A HIRES spectrum obtained by
one of us (JXP) also exhibits \ion{Si}{i} and \ion{Fe}{i}, ions also
reported by D'Odorico (2007), although neither of these cases has
coverage of higher ionization species.  The \nhi\ of the $z_{\rm abs}
= 0.745$ absorber towards Q1331+17 is not known, but its $D$-index
(Ellison 2006; Ellison et al. 2008) indicates that it is likely to be
a sub-DLA.  There are thus three cases where sub-DLAs exhibit
absorption from neutral gas, presumably a cold phase, despite their
relatively low \nhi.  In at least two of these cases, higher
ionization species are also present, a strong indication that the
sightline is intersecting multi-phase gas.  Although it is difficult
to draw robust conclusions from such small numbers, one explanation
could be that the \nhi\ column density of \textit{some} sub-DLAs is
low because of ionization from \hi\ to \ion{H}{ii}, as well as a
conversion of \hi\ to H$_2$ (e.g. Schaye et al. 2001; Krumholz et
al. 2009).  Although Noterdaeme et al. (2008) do not find a strong
correlation between \nhi\ and the probability of H$_2$ detection, the
sample of the sub-DLA population is relatively sparse.  A more
extensive survey of neutral ions such as \ion{C}{i} in DLAs and
sub-DLAs would be most interesting in this regard.

\section*{Acknowledgments} 

SLE is supported by an NSERC Discovery Grant. The data presented herein were
obtained at the W.M. Keck Observatory, which is operated as a scientific
partnership among the California Institute of Technology, the University of
California and the National Aeronautics and Space Administration. The
Observatory was made possible by the generous financial support of the W.M. Keck
Foundation.  The authors wish to recognize and acknowledge the very significant
cultural role and reverence that the summit of Mauna Kea has always had within
the indigenous Hawaiian community.  We are most fortunate to have the
opportunity to conduct observations from this mountain. NM also wants to
acknowledge that as the University of Victoria affiliates we are visitors on the
traditional territory of the Coast Salish people and we thank the WS'ANEC'
(Saanich), Lkwungen (Songhees) and Wyomilth (Esquimalt) peoples for an
opportunity to work and reside in their lands.

\begin{appendix}
  
\section{Extended element study in Cloudy}

\begin{figure}
\centerline{\rotatebox{0}{\resizebox{8cm}{!}{\includegraphics{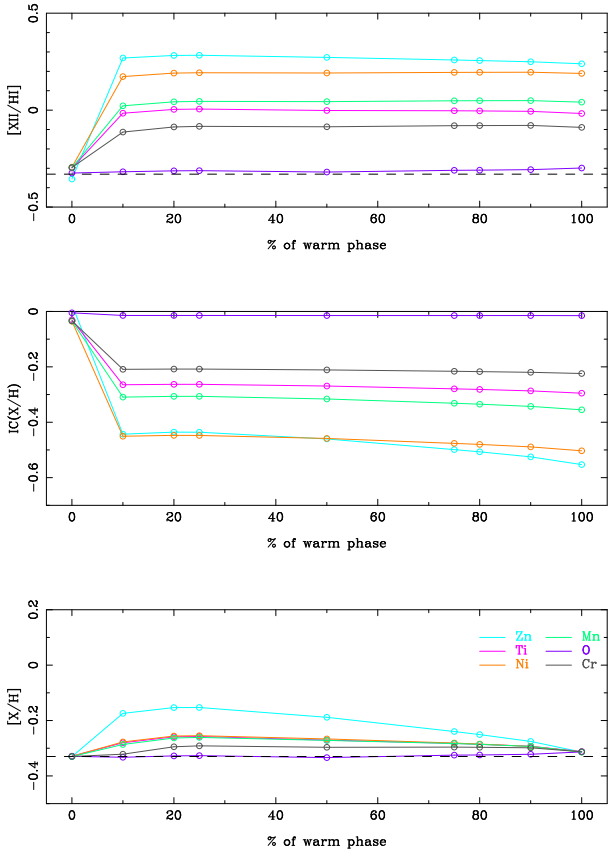}}}}
\caption{\label{ex1_full} Ionization corrections for the theoretically
modeled medium with [M/H]=$-0.33$. The ionic abundances `measured'
from the Cloudy models are presented in the top panel for different
elemental species commonly measured in DLAs and sub-DLAs. The
ionization corrections are given in the middle panel, and the
corrected abundances are shown in the bottom panel. The black
horizontal line represents the input metallicity of [M/H]$=-0.33$. The
single-phase based corrections recover the metallicity of the clouds
to within $\sim$0.15~dex.}
\end{figure}

\begin{figure}
\centerline{\rotatebox{0}{\resizebox{8cm}{!}{\includegraphics{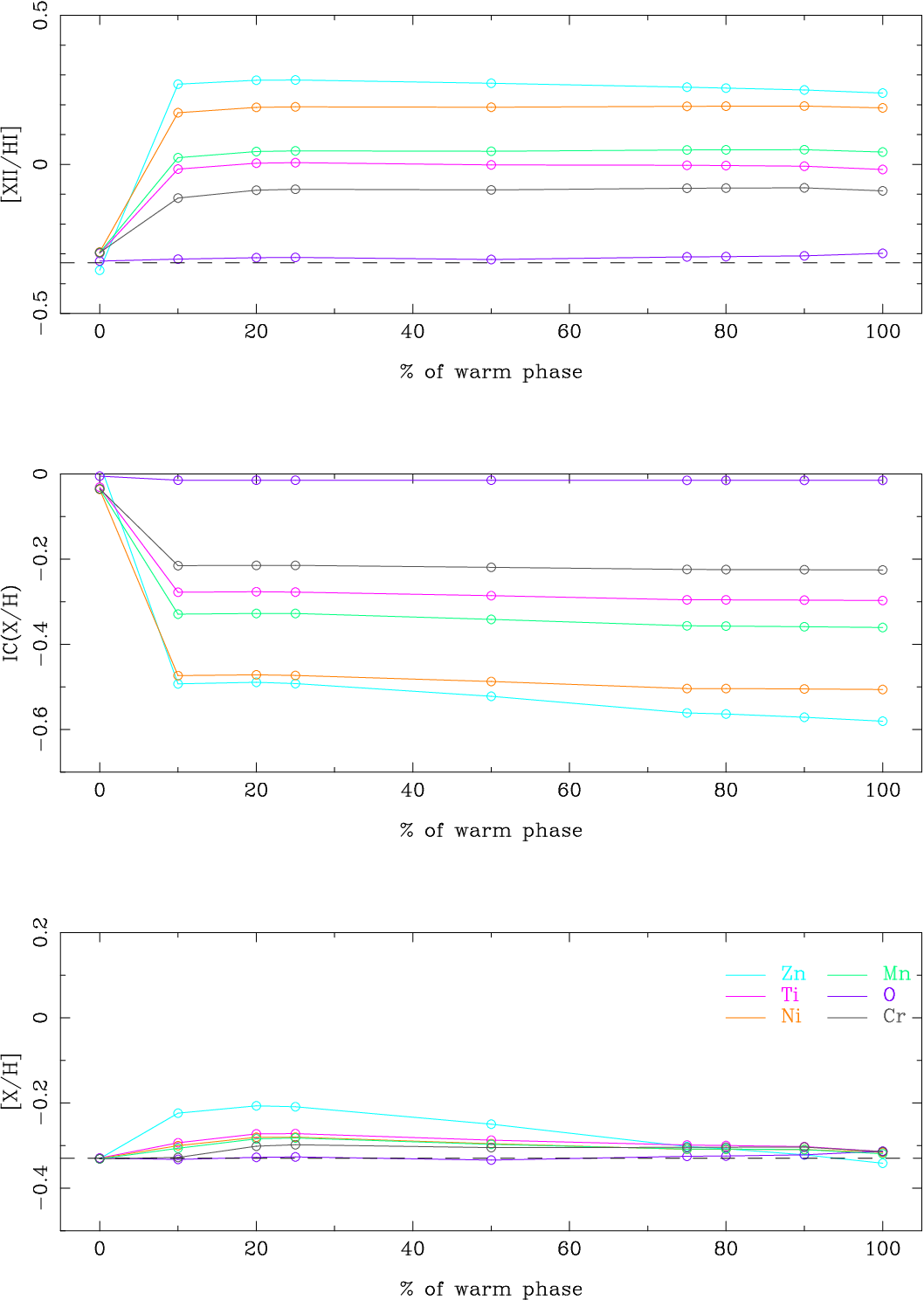}}}}
\caption{\label{ex2_full} Same as in Figure~\ref{ex1_full}, but for an
assumed metallicity of [M/H]=[\ion{Fe}{ii}/\ion{H}{i}].}
 \end{figure}

Figures \ref{ex1_full} and \ref{ex2_full} show the results of
the Cloudy modeling described in Section \ref{cloudy_sec} for a further
six elements commonly observed in DLAs.  These are presented here
for completeness, but we note that these corrections are not
generally applicable to the sub-DLA or DLA population as they
are modelled for the particular parameters of J2123$-$0050.
	
\end{appendix}

\end{document}